# Frequency dependence of dielectric anomaly around Neel temperature in bilayer manganite $Pr(Sr_{0.1}Ca_{0.9})_2Mn_2O_7$

Barnali Ghosh,[1] Dipten Bhattacharya,[2,*] A.K. Raychaudhuri,[1] and S. Arumugam[3]

[1]*Department of Materials Science, S.N. Bose National Center for Basic Sciences, Kolkata 700098, India*
[2]*Sensor & Actuator Division, Central Glass and Ceramic Research Institute, Kolkata 700032, India*
[3]*Department of Physics, Bharathidasan University, Tiruchirapalli, India*

(27 February, 2009)

A novel frequency dependence of anomaly in dielectric constant versus temperature plot, around the Neel temperature $T_N$ (~150 K), has been observed in a single crystal of bilayer manganite $Pr(Sr_{0.1}Ca_{0.9})_2Mn_2O_7$. The anomaly in the permittivity ($\varepsilon' \| c$) occurs at a temperature $T_f$ which moves within a temperature window ($\Delta T_f$) of ~40 K around $T_N$ for a frequency range 50 kHz – 5 MHz. The capacitive component $C_p$ of the dielectric response exhibits a clear yet broad feature over ~$\Delta T_f$ around $T_N$ which establishes the intrinsic capacitive nature of the anomaly.

PACS Nos. 75.80.+q

---

[*]e-mail: dipten@cgcri.res.in

## I. INTRODUCTION

Since the observation of flop of polarization (*P*) under a magnetic field (*H*) in TbMnO$_3$,[1] the subject of multiferroicity has grown rapidly, during the last few years, generating interest both as a problem of basic physics as well as an area of new application potential. Unconventional mechanisms such as incommensurate magnetic order (especially, spiral) driven breakdown of symmetry,[2] interface strain between ferroelectric and magnetic phases in composite or multilayer superlattice structures,[3] charge/orbital order driven non-centrosymmetric Born effective charge distribution[4-7] etc seem to be instrumental in offering strong coupling between the electric and magnetic order parameters in these multiferroics. The coupling yields a sizable magnetodielectric effect which ranges from 5-25% in most of the systems to as high as ~500% in compounds such as DyMnO$_3$ [Ref. 8] and CdCr$_2$S$_4$ [Ref. 9]. Among other mechanisms of multiferroicity, the one[4-7] of charge and orbital order (CO-OO) led non-centrosymmetry in Born effective charge distribution and consequent local polarization [as observed in LuFe$_2$O$_4$, Pr$_{1-x}$Ca$_x$MnO$_3$ (x = 0.4-0.5) etc] occupies a special position. This has given rise to a new class of multiferroics, where non-centrosymmetry arises from electronic origin, and can be termed as "electronic multiferroics".

It has been noticed that most of the "single phase" multiferroic compounds, however, exhibit a strong dielectric relaxation within the ac field frequency and temperature window of the experiments.[10] In such cases, the dielectric anomaly near magnetic transition point (*T$_N$*) is associated with strong relaxation feature – shift toward higher temperature with the increase in frequency. And the magnetodielectric effect originates mostly from the magnetoresistive response, in those cases, and not from the change in the dielectric constant.

In this backdrop, we report here observation of a genuinely "*frequency-dependent multiferroicity*" – shift in the anomaly in intrinsic dielectric constant over a temperature window of $\Delta T_f \sim 40$ K around $T_N$ (~150 K) in a single crystal of bilayer manganite $Pr(Sr_{0.1}Ca_{0.9})_2Mn_2O_7$ compound. This system enters into a charge/orbital ordered state below $T_{CO1}$ (~370 K) where charge/orbital order organizes into a stripe phase arrangement.[11,12] The stripes undergo a 90° rotation within the crystallographic ab-plane below a second transition point $T_{CO2}$ (~300 K).[11] This rotation is claimed to be responsible for developing a polar structure exhibiting long-range ferroelectricity. Therefore, this system is expected to exhibit a linear coupling between the electric and magnetic order parameters below the magnetic transition point $T_N$. We investigated the frequency-dependent multiferroicity in a high quality single crystal of $Pr(Sr_{0.1}Ca_{0.9})_2Mn_2O_7$ in detail.

**II. EXPERIMENTS**

The experiments have been carried out on a good quality single crystal of $Pr(Sr_{0.1}Ca_{0.9})_2Mn_2O_7$ prepared by traveling solvent zone method in an image furnace. The dc magnetization (*M*) versus temperature (*T*) pattern has been measured over a temperature range 2-400 K. A dc magnetic field (*H* = 100 Oe) was applied parallel to the c-axis. The zero-field dielectric properties have been measured on a sample of diameter 5 mm and thickness 0.5 mm, using parallel plate capacitor configuration, over a temperature range 90-220 K across the magnetic transition point $T_N$ (~150 K) for a frequency range 100 Hz – 5 MHz. Silver electrodes have been used for the measurements. The ac electric field ($E \sim 1$ V) was applied parallel to the c-axis.

### III. RESULTS AND DISCUSSION

The $M$-$T$ curve in Fig. 1 shows the expected features at $T_{CO1}$ and $T_{CO2}$ and reveals the $T_N \approx 150$ K. The pattern is consistent with that observed earlier for $H \parallel c$.[11] In Fig. 2, we show the real and imaginary parts of the permittivity $\varepsilon'(\omega,T)$, $\varepsilon''(\omega,T)$ across $T_N$. Clear anomaly near $T_N$, with frequency-dependent shift toward higher temperature, could be noticed. The shift takes place within a window of ~40 K around $T_N$ for a frequency range 50 kHz – 5 MHz (Fig. 2 inset). One important distinction between the observation made here and the ones made[10,13] in many other multiferroic systems (e.g., in $TbMnO_3$, $DyMnO_3$, $GdMnO_3$, $BiFeO_3$ etc) is the presence of strong dielectric relaxation along with anomaly arising out of coupling between electric and magnetic order parameters. The dielectric constant in those cases[10,13] is sufficiently high (>$10^3$) within the temperature and frequency window of interest and cannot be genuinely intrinsic. In the present case, the dielectric constant versus temperature plots for different frequencies across 50 kHz – 5 MHz nearly merge with each other or vary within a very narrow band and the dielectric constant tends to a frequency and temperature independent universal value of ~25 away from the temperature zone $\Delta T_f$ both at the low and high temperature ends. Moreover, the sharp nature of the anomaly itself signifies that its origin is associated with phase transition such as the onset of antiferromagnetic order. Therefore, the frequency-dependent shift in the dielectric anomaly within ~40 K around $T_N$ is a reflection of genuinely frequency-dependent multiferroicity where relaxed and truly intrinsic dielectric response exhibits the effect. This is an important result of this paper.

In order to probe this effect further, we examined the impedance spectroscopy over the temperature range 100-200 K. In Fig. 3, we show the data in the complex $Z'$-$Z''$ plane for a few temperatures at below and within $\Delta T_f$. The equivalent circuit (Fig. 3a inset) analysis of the spectroscopy shows that the dielectric anomaly originates primarily from the intrinsic capacitive mechanism while the dielectric relaxation originates from intrinsic response of the crystal with no contribution from the electrode-sample interface. The interface, therefore, is Ohmic and does not offer Maxwell-Wagner capacitance within the temperature range of interest. We used the generalized Davidson-Cole type relaxation model

$$Z^* = R_\infty + \frac{R_0 - R_\infty}{(1+j\omega\tau)^\beta} \qquad (1)$$

for fitting the experimental data; $R_0$ and $R_\infty$ are the static and high frequency resistances, respectively; $\omega$ is the frequency and $\beta$ designates the Kohlrausch exponent which is 1 for pure Debye relaxation and varies within 0 to 1 for non-Debye relaxation. The circuit parameters $R_p$, $C_p$, and relaxation time scale $\tau$ have been estimated and are plotted as functions of temperature ($T$) in Fig. 4. The relaxation, of course, follows non-Debye pattern with variation in the Kohlrausch parameter $\beta$ within 0.75-1.0 over the entire temperature range. The variation in $\beta$ with temperature is shown in Fig. 4b; $\beta$ tends to saturate at ~0.75 at below ~130 K and starts rising toward ~1.0 at higher temperature (130-200 K). The onset point of the saturation in $\beta$ matches closely with the peak point in the capacitive component $C_p$, extracted from the equivalent circuit analysis. It is interesting to note that there is a broad yet clear anomaly in intrinsic capacitance $C_p$ around $T_N$ which establishes unambiguously the capacitive origin of the dielectric anomaly observed. The resistive part $R_p$ does not show any visible anomaly near $T_N$. The $R_p$-$T$ pattern follows small polaron hopping transport model both at below and above $T_N$: $R_p = R_{p0}T\exp(E_{AR}/k_BT)$ and, as expected, $\tau$-$T$ follows Arrehenius model reasonably well, $\tau = \tau_0\exp(E_{AC}/k_BT)$. The activation energies $E_{AR}$ and $E_{AC}$ are found to be ~0.175 eV and ~0.15 eV at below $T_N$ and ~0.055 eV and ~0.283 eV at above $T_N$, respectively. The large difference between $E_{AR}$ and $E_{AC}$ at above $T_N$ signifies that the dielectric relaxation does not originate from small polaron hopping transport (i.e., from resistive mechanism) in this temperature regime.[14] At below $T_N$, of course, the transport is probably greatly influenced by the magnetic order due to the formation of magnons (spin-polarons) which, in turn, follows closely the fluctuation dynamics of polar charge/orbital stripe domains as a result of strong spin-charge-orbital coupling. Therefore, both the relaxation dynamics of polar domains and charge transport dynamics exhibit similar activation energy values.

This investigation reveals that the intrinsic capacitance ($C_p$) does exhibit a clear yet broad anomaly around $T_N$. In fact, the $C_p$-$T$ pattern closely resembles the $\varepsilon'(\omega,T)$-$T$ pattern. The role of the resistive component is conspicuously small. It establishes the fact that there is a sizable multiferroicity in this bilayer manganite Pr(Sr$_{0.1}$Ca$_{0.9}$)$_2$Mn$_2$O$_7$ which has not been reported so far. The broad anomaly in $C_p$-$T$ around $T_N$ might *also* give rise to the frequency-dependent anomaly in dielectric constant versus temperature plot within $\Delta T_f$. Such an interesting frequency dependence of anomaly in intrinsic dielectric constant around the magnetic transition point has not been observed in any other multiferroic system so far. In order to understand the origin of this effect, we measured the ac susceptibility across $T_N$ (data not shown here). We see no frequency-dependent shift in $T_N$ in the ac susceptibility. The magnetic order appears to be truly long-range and no signature of spin-glass-like behavior could be noticed. The frequency-dependence of the dielectric anomaly, therefore, possibly originates from a fluctuation in the coupling dynamics between polar domains of orbital stripes and long-range magnetic order. In other words, the coupling parameter $K$ between $\bm{P}$ and $\bm{M}$ [$\mathcal{H}(\omega) = K(\omega)\bm{P}.\bm{M}$] could turn out to be frequency dependent which could be manipulated by changing the frequency ($f$) of the ac electric field. The exponential growth of $T_f$ with $f$ (Fig. 2 inset) possibly signifies an exponential growth in the coupling parameter $K(\omega) \sim \exp(\omega)$. More detailed work, however, is needed to directly image the electric and magnetic domains and their coupling dynamics under varying field and frequency.

## IV. SUMMARY

In summary, we report observation of frequency-dependent shift in the dielectric anomaly near the magnetic transition point $T_N$ in a bilayer manganite Pr(Sr$_{0.1}$Ca$_{0.9}$)$_2$Mn$_2$O$_7$: perhaps an electronic multiferroic system where ferroelectricity originates from rotation of charge/orbital stripes (an electronic phase) with respect to the crystallographic structure and *not* from any crystallographic transition or soft phonon modes. The anomaly originates primarily from the capacitive component of the overall dielectric response ruling out the dominance of the resistive effect. This frequency-dependent effect could originate from fluctuating localized polar domains with locally switchable coupling.

**ACKNOWLEDGMENTS**

One of the authors (DB) thankfully acknowledges help from P. Mondal and S. Goswami during the experiments. BG thanks DST for the project under Women Scientist Scheme while AKR thanks DST for a sponsored project.

**Figure Captions**

**Fig. 1.** Dc magnetization versus temperature plot showing the charge/orbital order and magnetic transition points.

**Fig. 2.** (color online) (a) Real and (b) imaginary parts of the permittivity $\varepsilon'(\omega,T)$, $\varepsilon''(\omega,T)$ versus temperature patterns for a few frequencies. The shift in the dielectric anomaly toward higher temperature with frequency ($f$) around $T_N$ (~150 K) is clearly evident. The shaded region defines the $\Delta T_f$; Inset (a): the $T_f$-$f$ pattern is shown.

**Fig. 3.** (color online) The complex plane impedance plots for different temperatures at (a) below the transition zone ($\Delta T_f$) and (b) at within the zone; the temperatures corresponding to the plots in top plate are – from outer arc to inner ones – 113, 115, 117, 119, 121, 123, 125, 127, and 129 K and the temperatures corresponding to the plots in bottom plate are – 131, 134, 137, 140, 143, 146, 149, 152, 155, 158, 160 K; the solid lines represent the fit with Davidson-Cole relaxation equation which yields the equivalent circuit parameters – $R_p$ and $C_p$ – and the Kohlraush parameter $\beta$. Inset (a) shows the equivalent circuit with circuit elements.

**Fig. 4.** (a) The equivalent circuit resistance ($R_p$) and capacitance ($C_p$) versus temperature plots across $T_N$; (b) the variation of the Kohlrausch parameter $\beta$ with temperature and (c) the dielectric relaxation time constant ($\tau$) versus inverse temperature pattern.

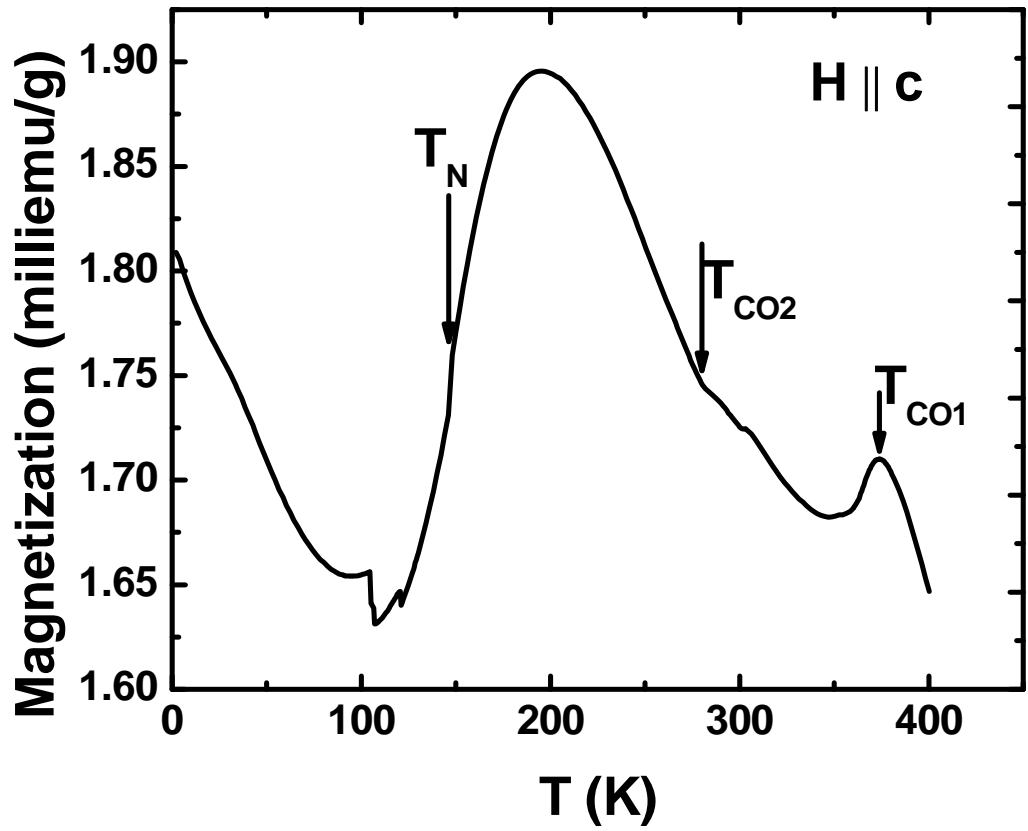

**Fig. 1.**

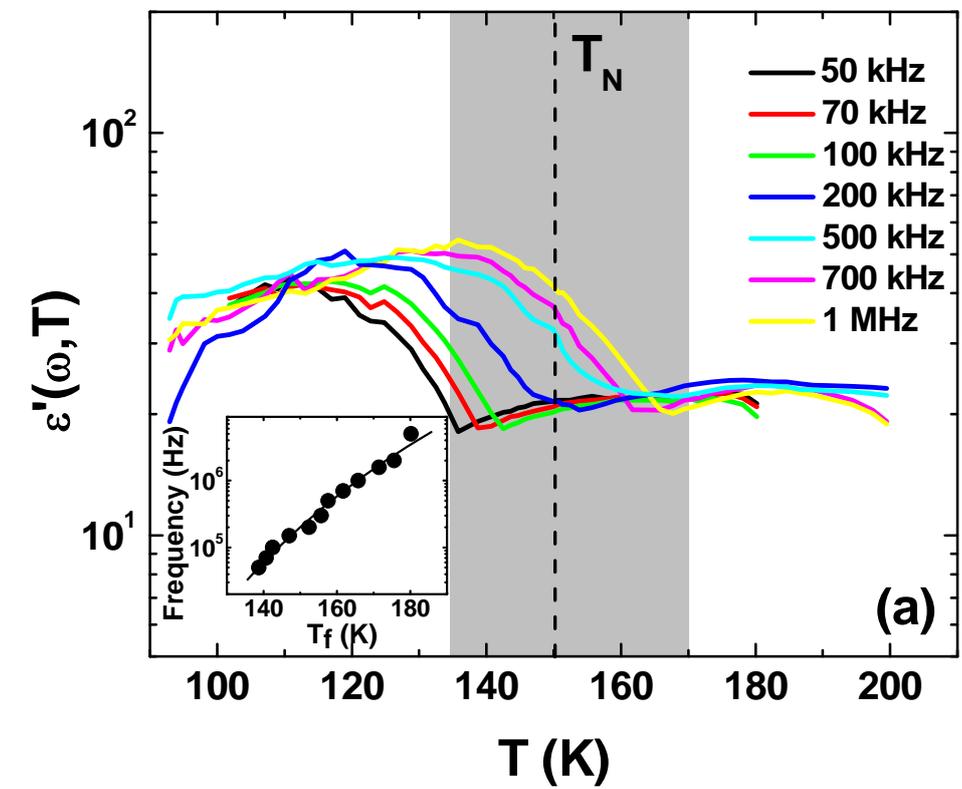
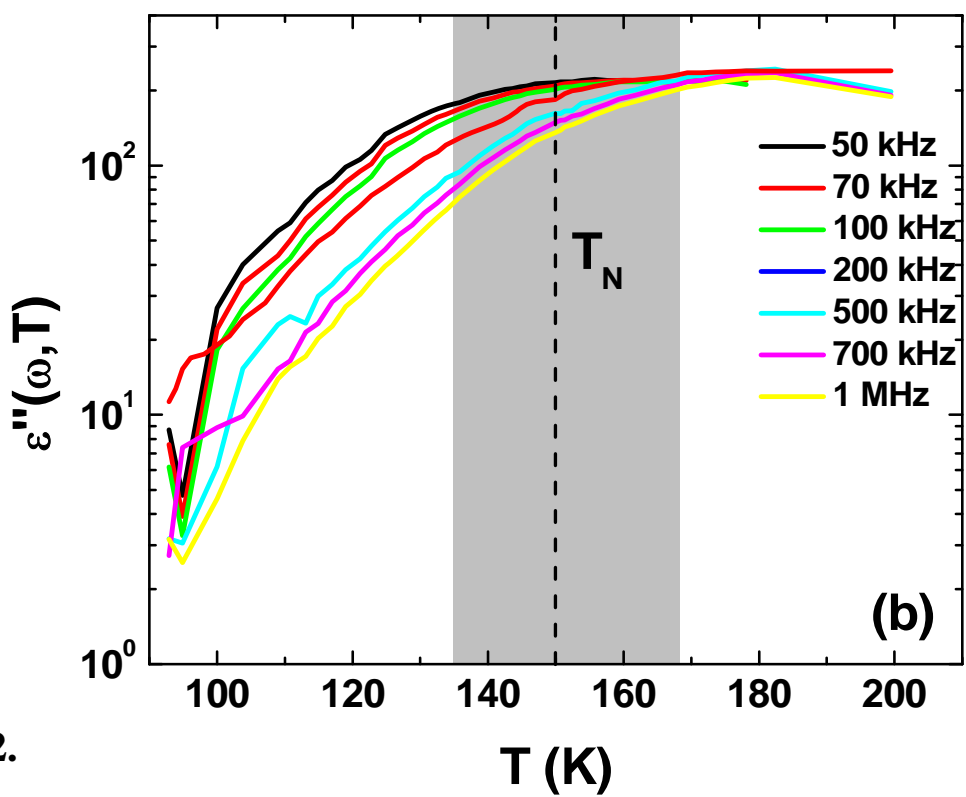

Fig. 2.

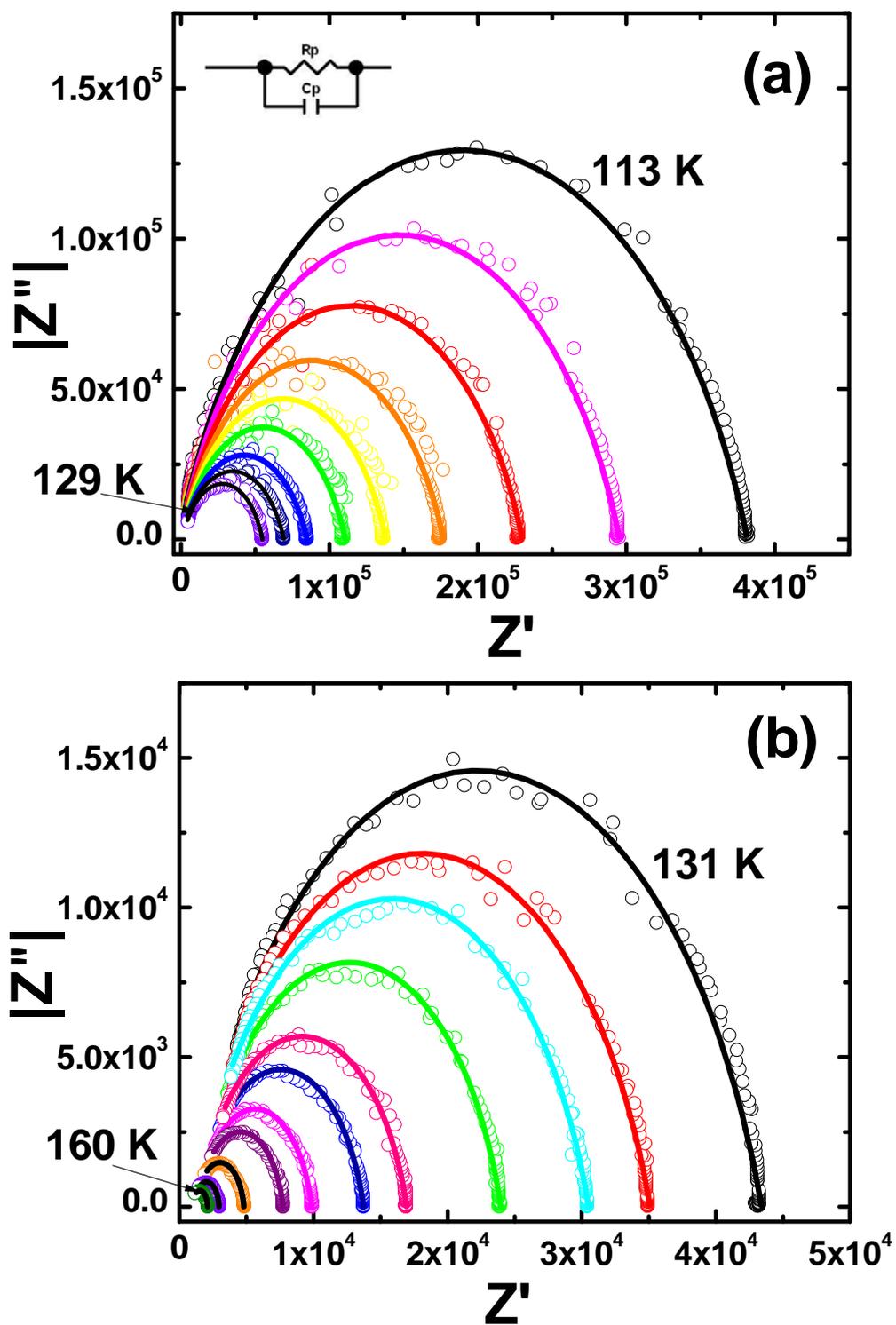

Fig. 3.

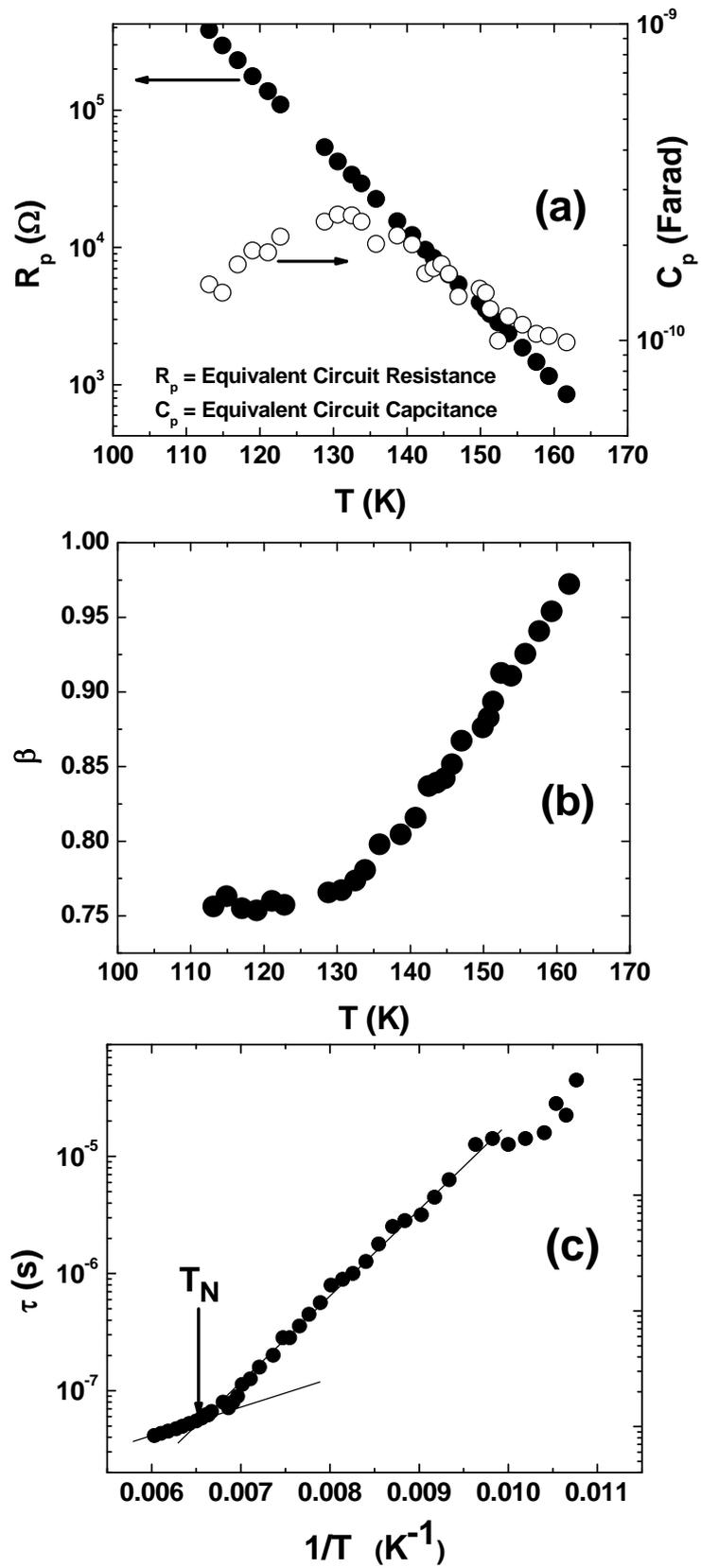

**Fig. 4.**